\documentclass[11pt,a4paper]{article}
\pdfoutput=1
\usepackage{jheppub}
\usepackage{graphicx,amssymb,amsmath,amsfonts}
\usepackage[english]{babel}
\usepackage{slashed}
\usepackage{multirow}
\usepackage{color}
\usepackage{braket}
\usepackage{simplewick}
\usepackage{pdfpages}
\usepackage{caption}
\usepackage{subcaption}
\usepackage{cleveref}
\usepackage{enumitem}

\usepackage{tikz}
\usetikzlibrary{decorations.pathmorphing}

\usetikzlibrary{arrows,positioning}

\def\beq{\begin{equation}}
\newcommand{\eeq}[1]{\label{#1}\end{equation}}
\def\bea{\begin{eqnarray}}
\newcommand{\eea}[1]{\label{#1}\end{eqnarray}}

\def\mS{\mathcal{S}}

\def\mO{\mathcal{O}}

\def\mS{\mathcal{S}}

\def\mH{\mathcal{H}}

\def\mI{\mathcal{I}}

\def\jb{\bar{j}}

\def\zb{\bar{z}}
\def\wb{\bar{w}}

\def\pa{\partial}

\def\g5{\gamma_5}

\def\b[#1]{\bold{#1}}
\def\bb[#1]{\overline{\bold{#1}}}

\def\bs[#1,#2]{\bold{#1}_{#2}}
\def\bbs[#1,#2]{\overline{\bold{#1}}_{#2}}

\def\s2{\sigma_2}

\def\ep{\epsilon}

\def\gammaflat{ \gamma_{z\zb}}
\def\gammaflatt{ \gamma^{z\zb}}

\def\gammaflatw{ \gamma_{w\wb}}
\def\gammaflatwt{ \gamma^{w\wb}}

\def\Rfp{R_f({p})}

\def\omegap{\omega_p}

\def\ketd[#1]{\ket{#1}_{\text{dressed}}}
\def\brad[#1]{\bra{#1}_{\text{dressed}}}

\def\ketas[#1]{\ket{#1}_{\text{Asymptotic}}}
\def\braas[#1]{\bra{#1}_{\text{Asymptotic}}}

\def\phiin{\phi_{\text{in}}}
\def\phiout{\phi_{\text{out}}}
\def\phiinhat{\hat{\phi}_{\text{in}}}
\def\phiouthat{\hat{\phi}_{\text{out}}}

\usepackage{etoolbox}% http://ctan.org/pkg/etoolbox
    \makeatletter
    \patchcmd{\maketitle}{\@fpheader}{}{}{}
\makeatother

\title{Comments on Lorentz Transformations, Dressed Asymptotic States and Hawking Radiation}
\author{Reza Javadinezhad$^a$, Uri Kol$^a$ and Massimo Porrati$^{a\,b}$}

\affiliation{a) Center for Cosmology and Particle Physics\\
	Department of Physics, New York University \\
	726 Broadway, New York, NY 10003, USA}
\affiliation{b) Center for Gravitational Physics\\ Yukawa Institute for Theoretical Physics\\ Kyoto University, Kyoto 606-8502, Japan\footnote{June 1 - August 31, 2018}}
\emailAdd{Rj1154,urikol,massimo.porrati@nyu.edu}

\abstract{

We consider two applications of the factorization of infrared dynamics in QED and gravity. The first is a redefinition of the 
Lorentz transformations that makes them commute with supertranslations. The other is the process of particle creation near a black hole horizon. For the latter we show that the emission of soft particles factors out of the S-matrix in the fixed-background approximation and to leading order in the soft limit. The factorization is implemented by dressing the incoming and outgoing asymptotic states with clouds of soft photons and soft gravitons.
We find that while the soft photon cloud has no effect, the soft graviton cloud induces a phase shift in the Bogolyubov coefficients relating the incoming and outgoing modes.
However, the flux of outgoing particles, given by the absolute value of the Bogolyubov coefficient, is insensitive to this phase. 
}

\begin{document}
\hfill YITP-18-86 
\maketitle

%%%%%%%%%%%%%%%%%%%%%%%%%%%%%%%%%%%%%%%%%%%%%%%%%%%%%%%%%%%%%%%%
%%%%%%%%%%%%%%%%%%%%%%%%%%%%%%%%%%%%%%%%%%%%%%%%%%%%%%%%%%%%%%%%
\section{Introduction}
%%%%%%%%%%%%%%%%%%%%%%%%%%%%%%%%%%%%%%%%%%%%%%%%%%%%%%%%%%%%%%%%
%%%%%%%%%%%%%%%%%%%%%%%%%%%%%%%%%%%%%%%%%%%%%%%%%%%%%%%%%%%%%%%%
Exclusive S-matrix elements in electrodynamics and gravity vanish because of the celebrated infrared ``catastrophe.'' 
One solution to the problem is to consider as physical only infrared-safe, inclusive cross 
sections~\cite{Bloch:1937pw,Kinoshita:1962ur,Lee:1964is}; the other is to ``dress'' asymptotic states by defining an 
intertwiner to a new Hilbert space, that maps Fock space states to non-Fock space ones. Physically, this 
means to multiply a state containing a finite number of particles with an operator 
that creates a coherent state with an infinite average number of soft 
photons/gravitons. The S-matrix between the dressed states has been argued to be free from (leading, soft) infrared 
divergences~\cite{Chung:1965zza,kibble1,Kibble:1969ip,Kibble:1969ep,Kibble:1969kd,Kulish:1970ut}.  Much more recently,
infrared degrees of freedom have been argued to be essential for a complete description of electrodynamics and gravity 
(see~\cite{Strominger:2017zoo} and references therein). It has been further argued to be a resource, rather than a problem, 
that can produce entanglement between (visible) hard particles, such as Hawking particles emitted by an evaporating 
black hole, and unobserved soft quanta~\cite{Strominger:2017aeh,Carney:2017jut,Carney:2017oxp}. 

This possibility is to be contrasted with the results of~\cite{Mirbabayi:2016axw,Bousso:2017dny,Bousso:2017rsx}, who argue
that dressed states and their associated dressed operators remove {\em in a universal and operationally simple way}
 the entanglement of hard particles with soft quanta.  Black holes and soft quanta have been studied in, among others, 
 refs.~\cite{Hawking:2016msc,Hawking:2016sgy,Averin:2016hhm,Gomez:2017ioy}.
 
 This paper will not delve into the deeper question of whether entanglement with soft particles can either solve or at least alleviate
 the black hole information problem. It will instead show how to use the infrared factorization of dynamics induced by the
  infrared dressing to simplify the dynamics of the
 radiative degrees of freedom. In particular, we will show that when the infrared dressing is applied to {\em operators} instead 
 of states, the resulting canonical transformation achieves two noteworthy effects. 
 
 The first is that one can define ``dressed'' deformed Lorentz transformations that  commute with both the  
 supertranslations charges and the boundary gravitons. The latter are  canonically conjugate to supertranslation charges. 
 After constructing these transformations and showing that their definition is consistent, we
  will briefly discuss the additional steps necessary to use this result to define new Lorentz transformations acting
   {\em on Hilbert-space states}, that are isomorphic to the standard Lorentz transformations, are symmetries of the radiative 
   degrees of freedom, and act trivially on all vacua related by supertranslations. If the charges associated to such 
   transformations do exist, they could replace the standard ones. Because all vacua related by supertranslations would be annihilated by the new charges, the latter would define unambiguously the angular momentum of
   any state. The standard charges instead give different outcomes even when evaluated on states that differ only by the 
   state of the soft quanta. We will define the new Lorentz charges in Section 2.
   
   Section 3 takes a detour and extend the construction of the dressing operator given in ref.~\cite{Gabai:2016kuf,Kapec:2017tkm} to the case when the total hard charge of
   an asymptotic state is nonzero. Section 3 establishes a bridge between the language 
   of~\cite{Chung:1965zza,kibble1,Kibble:1969ip,Kibble:1969ep,Kibble:1969kd,Kulish:1970ut,Gabai:2016kuf,Kapec:2017tkm} and that 
   of~\cite{Bousso:2017dny}. 
   
   Section 4 applies the formalism developed in this paper, that is the construction of dressed operators
    defined by the canonical
   transformation generated by the dressing operator, to the computation of Hawking's radiation in the limit $G \rightarrow 0$,
   $2GM=R_S=\mbox{constant}$. The fixed Schwarzschild radius $R_S$ and the vanishing Newton's constant $G$ eliminate 
   back reaction effects --and also guarantee that our calculation will not resolve one way or another the diatribe on 
   entanglement of Hawking quanta with soft quanta. On a positive note, this approximation will allow us to take into account exactly the 
   entanglement of the Hawking radiation with zero-energy quanta. We find that entanglement with photons is trivial, while
   entanglement with soft gravitons induces a 
   phase shift in the Bogolyubov coefficients relating the incoming and outgoing modes.
However, the flux of outgoing particles, given by the absolute value of the Bogolyubov coefficient, turns out to be
 insensitive to this phase. Local asymptotic observables defined at future null infinity are also insensitive to the phase.
We therefore conclude that the entanglement between soft and hard states due to infrared dynamics does not affect 
Hawking's radiation, when back-reaction is negligible. 

\subsection{Dressing as a canonical transformation}
Here we use the results of~\cite{Bousso:2017dny} to define a quantum canonical transformations of the operator algebra that
maps the standard asymptotic local fields into ``dressed'' fields that commute with the soft charges. We will be  general and rather
brief, deferring some explicit formulas to the next sections and referring to~\cite{Bousso:2017dny} for definitions not given in 
this paper. 
We will work in flat or asymptotically flat spacetimes; for such spaces the metric near future null infinity  
$\mathcal{I}^+$, is, in an appropriate gauge
\beq
 ds^2= -du^2 - 2dudr + r^2 \gamma_{AB} d\Theta^A d\Theta^B 
+ rC_{AB} d\Theta^A d\Theta^B +...
 \eeq{m1}
 The coordinates of $\mathcal{I}^+$ are two angles, $\Theta^A$ and the retarded time $u$. The 
 ``Bondi news'' tensor is $N_{AB}=\partial_u C_{AB}$; it is of course zero for Minkowski space. 
 The metric near past null infinity, $\mathcal{I}^-$ is defined in terms of angles $\theta^A$, radius $r$ and advanced time $v$:
\beq
 ds^2= -dv^2 - 2dvdr + r^2 \gamma_{AB} d\theta^A d\theta^B 
+ rC^-_{AB} d\theta^A d\theta^B +....
\eeq{m2} 
The past Bondi news is $N^-_{AB}=\partial_v C^-_{AB}$.  Both large gauge transformations (LGT) 
\cite{He:2014cra,Kapec:2015ena}  and BMS transformations~\cite{Bondi:1962px,Sachs:1962wk} are generated by
charges that can be defined, in the absence of massive matter, either at $\mathcal{I}^+$ or at $\mathcal{I}^-$. The superscript $\pm$ will distinguish between the two sets of charges. The charges are
functions of the two angles parametrizing the ``celestial sphere,'' i.e. the $S^2$ in $\mathcal{I}^\pm=\mathbb{R}\times S^2$. A
convenient decomposition of the charges is given
in terms of spherical harmonics, labelled as usual by the angular momentum $l$ 
and magnetic quantum number $m$. So we end up with two sets of charges, $Q^\pm_{lm}$. 
The equation expressing conservation of the total charges is very simple: once the angles $\Theta^A$ and $\theta^A$ 
are identified {\em antipodally} (see e.g.~\cite{Strominger:2017zoo}) then
the conservation equation is
\beq
  Q^+_{lm}=\Omega^{-1} Q^-_{lm} \Omega= Q^-_{lm}.
\eeq{m3}
Here $\Omega$ is the Heisenberg time evolution operator from $\mathcal{I}^-$ to $\mathcal{I}^+$. This equation says that 
$Q^\pm_{lm}$ commute with the evolution operator.
The charges $Q^\pm_{lm}$ are the sum of two pieces
\beq
 Q^\pm_{lm}= Q^\pm_{h\, lm}+ Q^\pm_{s\, lm}.
 \eeq{m4}
Let us discuss only 
the $+$ charges, as the $-$ are completely analogous. The ``hard'' charges $Q^+_{h\, lm}$ are defined only 
in terms of nonzero-frequency, a.k.a. radiative degrees of freedom.
For LGT these are the $U(1)$ field-strength components $F_{uA}$, plus eventual matter degrees of freedom. For BMS, the radiative degrees of freedom in the Bondi gauge
are the Bondi news $N_{AB}$ plus eventual matter degrees of freedom. 
The soft charges $Q^+_{s\,lm}$ commute with all the radiative degrees of freedom. The LGT soft charges are nonzero for 
$l>0$ while the BMS soft charges are nonzero for $l>1$.  In order to be dynamical, the soft charges need  conjugate variables
with nonzero commutators --otherwise they would be constants in each irreducible representation of the algebra of fields carrying radiative degrees of
freedom. The conjugate variables to the soft charges $Q^+_{s\,lm}$ are defined at $\mathcal{I}^+_-$, while those conjugate
to $Q^-_{s\,lm}$ are defined at $\mathcal{I}^-_+$. In the case of LGT they are sometimes called ``boundary photons'' 
while in the BMS case they are called ``boundary gravitons." In the rest of the Introduction we will call them  
$\Phi_{\pm\, lm}$ in both cases. 
Ref.~\cite{Bousso:2017dny} shows that for both LGT and BMS there exist a formally unitary transformation $U_\pm$ that
maps the soft charge into the total charge while it leaves the conjugate variable invariant
\beq
U_{\pm}Q^\pm_{s\,lm}U^{-1}_{\pm} = Q^\pm_{lm}, \qquad U_{\pm}\Phi_{\pm\, lm}U^{-1}_{\pm}=\Phi_{\pm\, lm} .
\eeq{m5}

The same operators $U_\pm$ defines canonical transformations that act on all observables of the theory. In the LGT case,
radiative variables of charge $q$ transform to ``dressed variables'' as 
\bea
\chi(u, \Theta)&\rightarrow &\hat{\chi} (u, \Theta) = U_+ \chi (u,\Theta)U^{-1}_+= e^{iq\Phi_+(u,\Theta)}\chi(u,\Theta) 
\nonumber \\
\chi(v, \theta)&\rightarrow &\hat{\chi} (v, \theta) = U_- \chi(v,\theta) U^{-1}_-= e^{iq\Phi_-(v,\theta)}\chi(v,\theta) ,
\eea{m6}
so the photon radiative modes are unaffected.
In the BMS case the transformation acts on all radiative degrees of freedom, including the Bondi news, as a translation (see \cite{Bousso:2017dny} and
Section \ref{BlackHole}). Denoting all these degrees of freedom collectively as $\phi$ we have
\bea
\phi^+(u, \Theta)&\rightarrow &\hat{\phi}^+ (u, \Theta) = U_+ \phi^+ (u,\Theta)U^{-1}_+ = \phi^+(u-\Phi_+(\Theta),\Theta) 
\nonumber \\
\phi^-(v, \theta)&\rightarrow &\hat{\phi}^- (v, \theta) = U_- \phi^-(v,\theta) U^{-1}_-= \phi^-(v-\Phi_-(\theta),\theta) .
\eea{m6}
The key property of the dressed (hatted) variables is that they commute with the {\em total} charges {\em and} with the conjugate variables. 
\bea
[\hat{\phi}^\pm , Q^\pm_{lm}] & = & [U_\pm \phi^\pm  U^{-1}_\pm , U_\pm Q^\pm_{s\, lm}  U^{-1}_\pm ]=
U_\pm \left[\phi^\pm, Q_{s\, lm}^\pm \right] U^{-1}_\pm=0, \nonumber  \\      
~[ \hat{\phi}^\pm , \Phi_{\pm\, lm} ]  &=&  [U_\pm \phi^\pm  U^{-1}_\pm , U_\pm \Phi_{\pm\, lm}  U^{-1}_\pm ]=
U_\pm [\phi^\pm, \Phi_{\pm\, lm} ] U^{-1}_\pm=0.
\eea{m7}
Besides the conservation law~\eqref{m3}, there exist also matching conditions for the boundary gravitons 
(and photons)~\cite{Strominger:2013jfa}
\beq
\Phi_{+\, lm}=\Omega^{-1}\Phi_{-\, lm}\Omega = \Phi_{-\, lm} .
\eeq{m8}

   The matching condition \eqref{m8} is imposed at $\mI^+_-$ and $\mI^-_+$. 
   In a geometry resulting from the collapse of matter with finite total energy causality guarantees that
   this asymptotic region is unaffected  by the    
   future evolution and ultimate fate of spacetimes. So it must hold for spacetimes that obey
   the Strong Asymptotic Flatness condition of Christodoulou and
   Klainermann~\cite{Christodoulou:1993uv} --even if they do not obey the global
   smallness assumption~\cite{Christodoulou:1993uv} so that they may
   contain black hole horizons. The charges $Q_{lm}^\pm$ were defined recently at
   spacelike infinity, where conservation is essentially ensured by
   definition~\cite{Hx1,Hx2,Hx3}; it is plausible that
   soft gravitons and photons could be likewise defined at spacelike infinity, where
   the matching condition~\eqref{m8} would be automatic.
   Moreover, \eqref{m8} is the only condition that leaves $\Phi_{\pm\, lm}$ invariant
   under CPT. Had we chosen different matching conditions, CPT transformations will map
   different super-selection sectors into each other and won't leave them invariant.
   Finally, the matching condition \eqref{m8} is the limit of the conservation of
   finitely-soft memories \cite{Bousso:2017rsx}, which is understood as the trivial
   propagation of soft degrees of freedom across spacetime to the antipodal angle.

Since the variables $\Phi_{\pm,\, lm}, Q^\pm_{lm}$ have canonical commutation relations among themselves {\em and
commute} with all the dressed variables, they can be represented canonically as 
$Q^\pm_{lm}, i \partial/\partial Q^\pm_{lm}$. Because of
 eqs.~(\ref{m3},\ref{m8}), the Heisenberg evolution operator then commutes with both $Q^\pm_{lm}$ and their derivatives, 
 so it depends only on the hatted radiative variables and is instead independent of the soft one. We conclude that the hatted 
 radiative  variables evolve independently of the soft variables as
 \beq
 \hat{\phi}^+=\Omega^{-1}(\hat{\phi}^-)\hat\phi^- \Omega(\hat\phi^-)=F[\hat\phi^-] .
 \eeq{m9}
 We should point out here that this derivation can be generalized to the case when in- and out- operators do not evolve 
 unitarily but are still related by a linear map. This is the case that applies to fields that scatter to $\mathcal{I}^+$ in a 
 collapsing black hole geometry. They are related to incoming fields defined on $\mathcal{I}^+$ by a linear map that becomes
 invertible (and is in fact 
 derived from a unitary evolution operator) only when the modes that cross the horizon of the black hole are
 taken into account. Yet, the linear map does not need the latter to be defined. This feature, that was prominently used
  in the original 1975 calculation of black hole radiation by Hawking~\cite{Hawking:1974sw,Hawking:1976ra}, will be used also in Section \ref{BlackHole}.

%%%%%%%%%%%%%%%%%%%%%%%%%%%%%%%%%%%%%%%%%%%%%%%%%%%%%%%%%%%%%%%%
%%%%%%%%%%%%%%%%%%%%%%%%%%%%%%%%%%%%%%%%%%%%%%%%%%%%%%%%%%%%%%%%
\section{Dressed Symmetry Group}
\subsection{Spacetime and its symmetries}
From now on we will work in the retarded coordinates $(u,r,z,\bar{z})$.
The metric of the flat spacetime in this coordinate is 
\begin{equation}\label{metric}
ds^2=-du^2-2du dr+2 r^2 \gamma_{z\bar{z}} dz d\bar{z} .
\end{equation}

In the above equation $\gamma_{z\bar{z}}=\frac{2}{(1+z\bar{z})^2}$ is the round metric on the unit 
sphere and $z=\tan(\frac{\theta}{2})e^{i\phi}$ is the coordinate defining the stereographic projection from the sphere 
to the complex plane. 
The metric (\ref{metric}) has ten Killing vectors, which we will use to find the representation of Poincar\'e generators on the 
phase space of the theory. We will consider first a $U(1)$ gauge theory and its associated LGT. Essentially, Poincar\'e transformation are induced by different components of the energy-momentum tensor. More precisely, the conserved charge 
\begin{align}\label{charge}
Q_\xi=\int_{\mathcal{I}^+} J^\nu d\Sigma_\nu=\int_{\mathcal{I}^+} 
\xi^\mu T_\mu^{~\nu} d\Sigma_\nu=\lim_{r\rightarrow\infty} \int_r \xi^\mu T_\mu^{~r} r^2\gamma_{z\bar{z}} dz d\bar{z}du, 
\end{align} 
induces transformation along the Killing vector $\xi$, when we use following equation 
\begin{align}\label{transf}
\delta_\xi\Phi=-i [Q_\xi,\Phi] .
\end{align}
\subsection{Gauge fixing and Lorentz transformations}
We work in radial gauge
\begin{equation}
A_r=0~,~A_u\vert_{I^+}=0 .
\end{equation}
A Lorentz transformation generically spoils the gauge condition; therefore, we will need a further 
gauge transformation to preserve the gauge condition (this can be thought of as a definition of the Lorentz generators). It can 
be shown that the extra gauge transformation only changes the leading order term of $A_u$ when expanded in powers of 
$r^{-1}$. It is convenient to write the explicit form of the leading order term of the gauge field near future null infinity in the 
following manner:
\begin{align}\label{asgauge}
A=\frac{A_u(u,z,\bar{z})}{r}du+A_z(u,z,\bar{z})dz+A_{\bar{z}}(u,z,\bar{z})d\bar{z} .
\end{align}
The E.O.M gives the following constraint on the gauge field
\begin{equation*}
A_u=\gamma^{z\bar{z}}(\partial_z A_{\bar{z}}+\partial_{\bar{z}} A_z)+g(z,\bar{z}) ,
\end{equation*}
where $g(z,\bar{z})$ is independent of $u$ but otherwise arbitrary function.
\subsection{Phase space construction and its symmetries}
The phase space of a $U(1)$ gauge theory plus massless particles near the 
asymptotic null infinities of an asymptotically flat spacetime is described by the set $\{F_{uz},F_{u\bar{z}},\Phi_+,\Psi_+ \}$, 
which consists of radiative degrees of freedom and soft degrees of freedom. Note that this phase space is defined
on $\mathcal{I}^+$; its canonical coordinates are functions of $u$ and of 
the coordinates on the celestial sphere $F_{uz}=F_{uz}(u,z,\bar{z}), \Phi_+=\Phi_+(z,\bar{z})$. 
\subsubsection{Lorentz transformation}
Next we find the representation of Lorentz generators acting on the phase space variables ($F_{uz},F_{u\bar{z}}$). Their gauge
potentials satisfy the  following commutation relation
\begin{align}
&\lbrace A_z(u,z,\bar{z}),A_{\bar{z}}(u',z',\bar{z}')\rbrace=-\frac{ie^2}{4}\Theta(u-u')\delta(z-z')\delta(\bar{z}-\bar{z}').
\end{align}
However, as we mentioned in the introduction, the pair ($F_{uz},F_{u\bar{z}}$)  is not enough to describe the complete
phase space, because we need two more fields $\Psi_+(z,\bar{z})$ and $\Phi_+(z,\bar{z}) $ defined at $\mathcal{I}^+_+$ 
and $\mathcal{I}^+_-$, respectively. The commutation relation for these new fields can be obtained from a ``continuity"
 condition on the commutators of the gauge potentials ($A_z,A_{\bar{z}}$).
It is simpler to introduce variables ($\tilde{\Psi}_+,\tilde{\Phi}_+$), related to ($\Psi_+,\Phi_+$) by a canonical 
transformation.
\begin{align}
&\tilde{\Phi}_+(z,\bar{z})\equiv\frac{\Psi_+(z,\bar{z})+\Phi_+(z,\bar{z})}{2} ,\\
&\tilde{\Psi}_+(z,\bar{z})\equiv \Psi_+(z,\bar{z})-\Phi_+(z,\bar{z}).
\end{align}
in terms of these fields, the commutation relations take the form
\begin{align}
&[\tilde{\Psi}_+(z,\bar{z}),A_{z}(u,w,\bar{w})]=\frac{ie^2}{4\pi}\frac{1}{z-w} ,\\
&[\tilde{\Psi}_+(z,\bar{z}),A_{\bar{z}}(u,w,\bar{w})]=\frac{ie^2}{4\pi}\frac{1}{\bar{z}-\bar{w}} ,\\
&[\tilde{\Phi}_+(z,\bar{z}),\tilde{\Psi}_+(w,\bar{w})]=-\frac{ie^2}{4\pi}\log\vert z-w\vert^2\label{commut} , \\
&[\tilde{\Phi}_+(z,\bar{z}),A_{z}(u,w,\bar{w})]=[\tilde{\Phi}_+(z,\bar{z}),A_{\bar{z}}(u,w,\bar{w})]=0 .
\end{align}
Equation (\ref{commut}) shows that $\tilde{\Phi}_+$ and $\tilde{\Psi}_+$ are conjugate variables. On the other hand, the soft charge is also conjugate to $\tilde{\Phi}_+$  (see e.g.~\cite{Bousso:2017dny}), so $\tilde{\Psi}_+$ and the soft charge cannot be independent variables. The soft charge is related to the boundary field $\tilde{\Psi}_+$ by following equation,
\begin{equation}
	Q^+_s[\delta^2(z-w))]=-\frac{2}{e^2} \partial_{\bar{w}}\partial_w \tilde{\Psi}_+(w,\bar{w}).
\end{equation}

An explicit form of the charge (\ref{charge}) can be written for this theory as
\begin{align}\label{excharge}
Q_\xi&= \frac{1}{e^2}\lim_{r\rightarrow \infty} \int_r \xi^\mu r^2 T_{\mu u}\gamma_{z\bar{z}} dz d\bar{z}du\nonumber\\&=
- \frac{1}{e^2}\int_{\mathcal{I}^+} [A_u(F_{uz}\xi^z+F_{u\bar{z}}\xi^{\bar{z}})+\gamma^{z\bar{z}}(2F_{uz}F_{u\bar{z}}\xi^u+F_{z\bar{z}}(F_{uz}\xi^z-F_{u\bar{z}}\xi^{\bar{z}}))]\gamma_{z\bar{z}} dz d\bar{z}du .
\end{align}  
The commutation relations of the conserved charge with the radiative and soft degrees of freedom are then given by
\begin{align}\label{az}
&[Q_\xi,F_{uz}(u',w,\bar{w})]=i\mathcal{L}_\xi F_{uz} ,\\
&[Q_\xi,F_{u\bar{z}}(u',w,\bar{w})]=i\mathcal{L}_\xi F_{u\bar{z}} ,\\
&[Q_\xi,\tilde{\Phi}^+]=0\neq i\mathcal{L}_\xi \tilde{\Phi}^+ ,\\
&[Q_\xi,\tilde{\Psi}^+]=i\mathcal{L}_\xi \tilde{\Psi}^+ .
\end{align} 
This shows that the boundary field $\tilde{\Phi}^+$ does not belong to a standard representation of Lorentz group, while $\tilde{\Psi}^+$ is in 
the scalar representation. This means that the states  of the $U(1)$ theory do not transform covariantly under Lorentz 
transformations --yet we can still have Lorentz invariant scattering amplitudes. In fact it has been known for a long time that 
there exists no Lorentz-covariant Hilbert space of charged physical states for gauge theories. Our motivation, however, is not
 to solve this problem. What we want instead 
 is to find a transformation that acts only on hard dressed variables and that commutes with
 their evolution operator. The factorization property of the S-matrix\footnote{See e.g. 
 ref~\cite{Mirbabayi:2016axw,Bousso:2017dny} for an approach close to the needs of this paper.} --or equivalently 
 the discussion in the Introduction--
implies the existence of an evolution operator that acts only on hard variables and leave soft variables untouched. We would 
like to have the same property for symmetry transformations, this is why we look for a symmetry transformation that acts 
nontrivially only on the hard states. 

More precisely, we want to change the standard definition of the Lorentz transformation in two ways.
First of all, we want to ``dress'' the Lorentz transformation with the operators $U_\pm$ defined in the Introduction.
In other words,  we want to perform the quantum canonical transformation 
\beq
Q_\xi \rightarrow \hat{Q}_\xi^\pm = U_\pm Q_\xi U^{-1}_\pm .
\eeq{mr1}
This transformation acts on the hatted, dressed operators exactly as the original Lorentz does on the undressed ones.
In the rest of the section we will work with dressed operators only, so can drop the hat from all fields to avoid notational clutter.
The second change to the Lorentz transformation is the crucial one: we change its action on the soft variables, while leaving
its action on the radiative variables unchanged. The Lorentz transformation is finally defined by the following commutators
\begin{align}\label{newl}
&[\tilde{Q}_\xi,F_{uz}(u',w,\bar{w})]=i\mathcal{L}_\xi F_{uz} ,\\
&[\tilde{Q}_\xi,F_{u\bar{z}}(u',w,\bar{w})]=i\mathcal{L}_\xi F_{u\bar{z}} ,\\
&[\tilde{Q}_\xi,\tilde{\Phi}^+]=0 ,\\
&[\tilde{Q^+}_\xi,\tilde{\Psi}^+]=0 .
\end{align} 
Evidently, this symmetry is not the usual one since it acts like Lorentz only on the hard variables. The
definition given above has to pass a crucial test of consistency: it must satisfy the Jacobi identities. 
Thanks to the fact that the new Lorentz transformation is the same as the standard one on radiative variables, it is easy to
  check that the Jacobi identities are indeed satisfied. The action of the symmetry on operators, given by eqs.~(\ref{newl}), 
  plus its action on the vacuum (which we can choose to be trivial) then {\em formally} define an operator acting on a
  Hilbert space. The Lorentz transformations defined here are automorphisms of the operator algebra. To show that 
  they are in fact generated by commutators of Hermitian operators, several other properties would have to be proven.
  The first is that the transformation maps a vector obtained by applying a polynomial function of dressed operators to one of 
  the vacua --i.e. a state containing only soft photons-- into a state belonging to the same Hilbert space. 
  The second is that such property holds for a dense set of states in the Hilbert space. These two properties define a linear 
  operator in the Hilbert space, that can be identified with the Lorentz charge only after a third property is proven; namely, that
  the domain of the operator and the domain of its adjoint coincide. In spite of these pitfalls, the definition given here provides
   a sketch of the construction of  a factorized ``dressed Lorentz,'' that can be thought of as a counterpart to the 
   factorization of the S-matrix. 

A similar construction can be done for gravity and BMS. Here we look for Lorentz transformations that leave the soft 
variables untouched and only change the hard degrees of freedom --which in this case are the Bondi news tensors. By 
denoting here with $C$  the boundary graviton the transformation laws that we look for are 
\bea
 ~[\tilde{Q}_\xi,N_{zz}] &=& i\mathcal{L}_\xi N_{zz}, \nonumber \\
 ~[\tilde{Q}_\xi,N_{\bar{z}\bar{z}}] &=&  i\mathcal{L}_\xi N_{\bar{z}\bar{z}}, \nonumber \\
 ~[\tilde{Q}_\xi,C] &=& 0, \nonumber \\
~[\tilde{Q}_\xi, Q_f] &=& 0.
\eea{newg}
Here the generators of supertranslation are denoted by $Q_f$. As before, these are transformations on dressed fields, where hats have been dropped.

The ``dressed Lorentz'' defined in~(\ref{newg})  commutes  with supertranslations, so it is an ideal subalgebra of BMS. We 
have to check again the Jacobi identity. It is easy to see that the only nontrivial equation to check is 
\beq
	[Q_f,[\tilde{Q}_\xi,C]]+[C,[Q_f,\tilde{Q}_\xi]]+[\tilde{Q}_\xi,[C,Q_f]]=0.
\eeq{mr2}
Since $Q_f$ and $C$ commute to a c-number, eq.~(\ref{mr2})  does satisfy
the Jacobi identity.

The undressed Lorentz transformations $Q_\xi$
  do not commute with supertranslations; this is true for boosts as well as
  rotations. Even on a state that contains no hard degrees of
  freedom supertranslations
  change the state of the soft hair. The non-commutativity of undressed rotations
 with supertranslations 
  then means that angular momentum depends on soft degrees of freedom and cannot
  be computed unambiguously from the knowledge of the state of hard degrees of
  freedom alone. The ambiguity in the definition of the angular momentum
  is an old problem classical general relativity~\cite{Corn}. If our definition of
  dressed Lorentz transformations can be promoted to a well-defined Hermitian
  operator acting on a Hilbert space, it will offer an interesting new
  definition of angular momentum free of ambiguities. Whether this new definition
  is operationally
  useful depends on an additional property that we did not and will not address
  in this paper; namely whether it can be expressed as an asymptotic
  boundary integral of
  local fields. This property, which is true for standard definitions of energy,
  momentum and angular momentum, allows the computation of such quantities
from local data evaluated in an asymptotic, weak gravity region.
%%%%%%%%%%%%%%%%%%%%%%%%%%%%%%%%%%%%%%%%%%%%%%%%%%%%%%%%%%%%%%%%
%%%%%%%%%%%%%%%%%%%%%%%%%%%%%%%%%%%%%%%%%%%%%%%%%%%%%%%%%%%%%%%%

\section{Asymptotic States}
%%%%%%%%%%%%%%%%%%%%%%%%%%%%%%%%%%%%%%%%%%%%%%%%%%%%%%%%%%%%%%%%
%%%%%%%%%%%%%%%%%%%%%%%%%%%%%%%%%%%%%%%%%%%%%%%%%%%%%%%%%%%%%%%%

Asymptotic states for Abelian gauge theories were studied in \cite{Chung:1965zza,kibble1,Kibble:1969ip,Kibble:1969ep,Kibble:1969kd,Kulish:1970ut}, and more 
recently in~\cite{Ware:2013zja,Choi:2017bna,Choi:2017ylo} for perturbative quantum gravity. In this section we derive 
covariant expressions for the dressing functions of the asymptotic states in terms of the fields at null infinity. We will 
concentrate on incoming asymptotic states at past null infinity, but to linear order in the fluctuations around a fixed 
background a similar 
analysis applies also to future null infinity, as well as 
to any other null surface. We generalize results known in the literature to the case where the total charge (gauge or 
supertranslations) is nonzero. We define the dressing function $\Rfp $ on one-particle states as
\begin{equation}
\ketas[\vec{p}] = e^{ \Rfp } \ket{\vec{p}}.
\end{equation}

\subsection{Abelian gauge theory}

We start with the case of Abelian gauge theory. The dressing function in this case is given by

\begin{equation}\label{abelianDress}
\Rfp=
-\frac{e}{(2\pi)^3}
\int \frac{d^3q}{2 \omega_q}
\left(f^{\mu}a_{\mu}^{\dagger}(\vec{q})-f^{*\mu}a_{\mu}(\vec{q})\right) ,
\end{equation}
where the function $f^{\mu}$ is given by
\begin{equation}
f^{\mu} = \left[\frac{p^{\mu}}{p\cdot q}-\frac{c^{\mu}(\hat{q})}{\omega_q}\right] \psi(p,q) ,\qquad q^\mu=\omega_q \hat{q}^\mu
\end{equation}
and $c^{\mu}$ satisfies $c\cdot \hat{q}=1$ and $c^2=0$. Notice that up to a conventional overall scale factor, the function 
$c^{\mu}=c^{\mu}(\hat{q})$ depends only on the celestial sphere's coordinates. 
An example of $c^{\mu}$ that satisfies the aforementioned condition is
 \begin{equation}
	c^{\mu} = \frac{1}{2(1+ w \wb)}\left(1+ w \wb,-w-\wb,i(w-\wb),-1+w \wb\right) .
	\end{equation}
While we will keep a generic $c^\mu$ in our formulas, this specific choice simplifes eqs.~\eqref{3.16}, \eqref{m-cloud} and particularly eq.~\eqref{3.22}.
$\psi(p,q)$ is a regulating smooth function that obeys $\psi(p,q)\rightarrow 1$ as $q \rightarrow 0$.

Next we write the dressing function in terms of the asymptotic gauge field at null infinity.
We use the following decomposition of the gauge field in terms of Fourier components
\begin{equation}
A_{\mu} (\vec{x}) = \int  \frac{d^3 q}{(2\pi )^3} \frac{1}{2\omega_q} \left[
a_{\mu}(\vec{k}) e^{i k \cdot x} +a_{\mu}^{\dagger} (\vec{k}) e^{-i k \cdot  x} 
\right] .
\end{equation}
The Fourier modes can be decomposed in terms of polarization modes
\begin{equation}
a_{\mu}(\vec{q}) = \ep ^{*\alpha}_{\mu}(\vec{q}) a_{\alpha}(\vec{q}), \qquad \alpha = \pm .
\end{equation}
We will focus on the form of the gauge field at past null infinity, where it takes the form
\begin{equation}
\begin{aligned}
A_w (v,w,\wb) &= \lim_{r \rightarrow \infty} A_w (v,r,w,\wb)  , \\
&= - \frac{i}{8 \pi ^2} \frac{\sqrt{2} e}{1+w \wb} 
\int _0^{\infty} d\omega_q \left[  a_+ (\vec{q}) e^{-i\omega_q v} - a_- ^{\dagger}(\vec{q}) e^{i\omega_q v}  \right] .
\end{aligned}
\end{equation}
The analysis for future null infinity (or any other null surface) can be carried out in a similar fashion, yielding similar results.
The reverse relations are given by
\begin{equation}
\begin{aligned}
a^{\dagger}_{+/-} (\vec{q}) &=
-\frac{4\pi i}{e\sqrt{\gammaflatw}}
\int dv e^{-i\omega_q v} A_{w / \wb} (v,w,\wb) ,\\
a_{-/+} (\vec{q}) &=
+\frac{4\pi i}{e \sqrt{\gammaflatw}}
\int dv e^{i\omega_q v} A_{w / \wb} (v,w,\wb) .
\end{aligned}
\end{equation}
Without loss of generality we write the regulating function in the following form
\begin{equation}
\psi(p,q) = e^{-i v_c \omega_q} ;
\end{equation}
we can then write the cloud operator \eqref{abelianDress} as\footnote{In terms of the advanced system of coordinates, the volume form is given by $d^3 q = \omega_q ^2 \gammaflatw d\omega_q d^2 w$. We use that $\frac{1}{2\pi} \int dv e^{-iv \omega} = \delta(v)$. Note that since the $\omega$-integration is over half the real plane we have $\int_0 ^{\infty} d \omega \delta(\omega) f(\omega) = \frac{1}{2} f(0)$.}
\begin{equation}
\Rfp =
\frac{iQ_0}{2\pi}
\int d^2 w \sqrt{\gammaflatw}
\left( \frac{p^{\mu}}{p\cdot \hat{q}} - c^{\mu} \right)
 \left(
  \ep^-_{\mu}  A_{w}(v_c,w,\wb)
 + \ep^+_{\mu}  A_{\wb}(v_c,w,\wb)
\right)  .
\end{equation}
Here $v_c$ is parametrizing the regularization scheme.

Since the photon is a massless particle its four-momentum can be written as
\begin{equation}
q^{\mu} = \frac{\omega_q}{1+ w \wb}\left(1+ w \wb,w+\wb,-i(w-\wb),1-w \wb\right) .
\end{equation}
For a massless external particle with energy $\omega_p$ located at $(z_0,\zb_0)$ on the celestial sphere the four-momentum is similarly given by
\begin{equation}
p^{\mu} = \frac{\omega_p}{1+ z_0 \zb_0}\left(1+ z_0 \zb_0,z_0+\zb_0,-i(z_0-\zb_0),1-z_0 \zb_0 \right) .
\end{equation}
The photon polarization vectors can be chosen to take the form
\begin{equation}\label{polarization}
\begin{aligned}
\ep^{+\mu}(\vec{q})&= \frac{1}{\sqrt{2}} \left(\wb,1,-i,-\wb\right)  ,\\
\ep^{-\mu}(\vec{q})&= \frac{1}{\sqrt{2}} \left(w,1,i,-w\right) ,
\end{aligned}
\end{equation}
such that they are orthogonal to $q^{\mu}$
\begin{equation}
q_{\mu} \ep^{\pm \mu}(\vec{q})=0, \qquad
\ep^{\mu}_{\alpha}\ep^{*}_{\beta \mu} = \delta_{\alpha \beta} .
\end{equation}
Using this parametrization one can evaluate
\begin{equation}
\begin{aligned}
\sqrt{\gammaflatw}\frac{p\cdot \ep^-}{p\cdot q} &=  \frac{1}{\wb  -  \zb_0}, \\
\sqrt{\gammaflatw}\frac{p\cdot \ep^+}{p\cdot q} &=  \frac{1}{w  -  z_0  }  .
\end{aligned}
\end{equation}
The expression for the cloud operator is then given by
\begin{equation}\label{3.16}
\begin{aligned}
\Rfp &=
\frac{iQ_0}{2\pi}
\int d^2 w 
\left[
\left( \frac{1}{\wb-\zb_0}- \sqrt{\gammaflatw} c \cdot \ep^- \right)   A_{w}(v_c,w,\wb)
\right.
\\
&  \left.
\qquad \qquad \qquad \qquad
+ \left( \frac{1}{w-z_0}   -   \sqrt{\gammaflatw} c \cdot \ep^+    \right)  A_{\wb}(v_c,w,\wb)
\right] .
\end{aligned}
\end{equation}
The Green's function for the radiative modes is given by
\begin{equation}
G(z,w) = \ln |z-w|^2
\end{equation}
and it obeys
\begin{equation}
\begin{aligned}
\pa_z \pa_{\zb} G(z,w) &= 2\pi \delta^{(2)}(z-w) ,\\
\pa_{z} G(z,w) & = \frac{1}{z- w}  ,\\
\pa_{\zb} G(z,w) & = \frac{1}{\zb- \wb} .
\end{aligned}
\end{equation}
The cloud operator can be expressed using the Green's function as follows
\bea
\Rfp&=&
\frac{iQ_0}{(2\pi)^2}
\int d^2 w \gammaflatw d^2z \gammaflat
G(z,w)
\left[
2\pi \gammaflatt  \delta^{(2)}(z-z_0 ) D\cdot A(v_c,w,\wb)
\right.
 \nonumber \\
&&  \left.  
\qquad \qquad \qquad \qquad\qquad\qquad
-  \pa^{\zb} \left(   \sqrt{\gammaflat} (c \cdot \ep^-)_z \right)   \pa^{w}  A_{w}(v_c,w,\wb)
\right.
\nonumber \\
&&  \left.  
\qquad \qquad \qquad \qquad\qquad\qquad
- \pa^{z} \left(     \sqrt{\gammaflat} (c \cdot \ep^+ )_z   \right) \pa^{\wb}  A_{\wb}(v_c,w,\wb)
\right] ,
\eea{m-cloud}
where $D\cdot A=D^{w}A_{w}+D^{\wb}A_{\wb}$.
In the above equation, the subscript on $(c \cdot \ep^{\pm})_z$ means that the expression inside the parentheses is evaluated at $(z,\zb)$, while previously it was a function of the position of the soft photons inside the cloud, $(w,\wb)$. For brevity, we will remove this subscript from now on, as the dependence on the coordinates is clear from the derivative acting on it.

At this point we would like to discuss the regularization scheme parametrized by $v_c$. Note that in order to isolate the contribution of zero momentum photons in the cloud we should consider $v_c = + \infty $ (see \cite{Bousso:2017dny,Bousso:2017rsx}). In the absence of magnetic monopoles and long-range magnetic fields
\begin{equation}
\pa_z A_{\zb} (+ \infty  , z,\zb)=\pa_{\zb} A_{z} (+ \infty  , z,\zb)
\end{equation}
The dressing operator then takes the final simplified form
\begin{equation}\label{photonCloud}
\Rfp=
\frac{i}{2\pi}
\int dv d^2 w \gammaflatw d^2z \gammaflat
\left[  j_v(v,z,\zb)-\bar{j}_v(v,z,\zb) \right]  G(z,w)   D\cdot   A(+ \infty ,w,\wb) ,
\end{equation} 
where
\begin{equation}\label{3.22}
\begin{aligned}
j_v(v,z,\zb) &=Q_0 \delta(v-v_0)  \gamma^{z\zb}  \delta^{(2)}(z-z_0 ) , \\
\bar{j}_v(v,z,\zb)& =\frac{Q_0}{2\pi} \delta(v-v_0) \gamma^{z\zb}  \left[  \pa_{z} \left(   \sqrt{\gammaflat} c \cdot \ep^- \right) +  \pa_{\zb} \left(     \sqrt{\gammaflat} c \cdot \ep^+   \right) \right] .
\end{aligned}
\end{equation}
Here $j_v(v,z,\zb)$ is the classical incoming gauge current and 
$\bar{j}_v(v,z,\zb)$ is interpreted as the LGT current, in accordance with ref.~\cite{Gabai:2016kuf}.
Eq.~\eqref{3.22} holds for generic $c^\mu$; the particular choice $c \cdot \ep^+=\frac{\bar{w}}{\sqrt{2}}$,$c \cdot \ep^-=\frac{w}{\sqrt{2}}$ and $\bar{j}_v(v,z,\zb) =\frac{Q_0}{\sqrt{2}\pi} \delta(v-v_0)$ makes the current $\bar{j}_v$ isotropic.

\subsection{Gravity}

We now turn to studying the dressing operator in perturbative quantum gravity, following 
refs.~\cite{Choi:2017bna,Choi:2017ylo}.
 It is given by
\begin{equation}\label{FKstateGravity}
\Rfp=
\kappa
\int \frac{d^3q}{(2\pi)^3 2\omega_q}
\left[   f^{\mu\nu*}a_{\mu\nu}^{\dagger}(\vec{q})-f^{\mu\nu}a_{\mu\nu}(\vec{q})\right], \qquad \kappa ^2 = 32 \pi G.
\end{equation}
The function $f^{\mu\nu}$ is given by
\begin{equation}
f^{\mu\nu} = \left[\frac{p^{\mu}p^{\nu}}{p\cdot q}-\frac{c^{\mu\nu}(\vec{p},\hat{q})}{\omega_q}\right] \psi(p,q) .
\end{equation}
As in the gauge theory case, we have factorized the magnitude $\omega_q$ from the function $c_{\mu\nu}$, and  $\psi(p,q)\rightarrow 1$ as $q \rightarrow 0$. For a detailed discussion about the function $c_{\mu\nu}$ and the constraints it should obey we refer the reader to \cite{Choi:2017bna}.

In a similar fashion to the previous subsection we would like to express the cloud operator in a covariant form, using the asymptotic form of the fields. We decompose the metric into Fourier components
\begin{equation}
h_{\mu\nu} (\vec{x}) = \int \frac{d^3 q}{(2\pi)^3} \frac{1}{\omega_q} \left[
a_{\mu\nu}(\vec{q}) e^{iq \cdot x}
+a^{\dagger}_{\mu\nu}(\vec{q}) e^{-iq \cdot x}
\right]  .
\end{equation}
Decomposing the Fourier components in terms of the polarization modes
\begin{equation}
a_{\mu\nu} = \sum_{r=\pm} \ep^{*r}_{\mu\nu}a_r ,
\end{equation}
the radiative data at past null infinity can then be written as
\begin{equation}
\begin{aligned}
C^-_{zz}(v,z,\zb) &= \kappa \lim_{r \rightarrow \infty}  \frac{1}{r} h_{zz}(r,v,z,\zb) \\
&=
- \frac{i \kappa}{8\pi ^2} \gammaflat \int_0^{\infty} d\omega_q  \left[a_+ (\vec{q})e^{-i\omega_q  v}-a_- (\vec{q})e^{+i\omega_q v}\right] .
\end{aligned}
\end{equation}
We can invert the expression to express the Fourier coefficients in term of the asymptotic fields
\begin{equation}
\begin{aligned}
a_{+ / -}(\vec{q}) &=
\frac{4\pi i }{\kappa \gammaflat} \int dv e^{+i \omega_q v} C^-_{\zb\zb / zz} (v,z,\zb) ,
\\
a^{\dagger}_{- / +}(\vec{q}) &=
-\frac{4\pi i }{\kappa \gammaflat} \int dv e^{-i \omega_q v} C^-_{\zb\zb / zz} (v,z,\zb) ,
\end{aligned}
\end{equation}
to write the cloud operator as
\begin{equation}
\Rfp = 
-\frac{i}{4\pi}
\int d^2 w
\omega_q \left[
(f\cdot \ep^+)C^-_{\wb\wb}(v_c,w,\wb)
+(f\cdot \ep^-)C^-_{ww}(v_c,w,\wb)
-\text{h.c.}
\right]  .
\end{equation}
Here $f\cdot \ep^{\pm} \equiv f^{\mu\nu}\ep_{\mu\nu}^{\pm}$, and we have used the 
regularizing function $\psi(p,q)=e^{-iv_c \omega_q}$, as in the Abelian gauge theory case.

The Green's function for the radiative modes obeys the equation
\begin{equation}
D^2 _z D^2 _{\zb} G(z,w) = 2\pi \gammaflat \delta^{(2)} (z-w) ,
\end{equation}
where $D$ is the covariant derivative on the 2-sphere (see, for example, \cite{Strominger:2014pwa,Campiglia:2015kxa,Campiglia:2015lxa}).
The equation above is solved by
\begin{equation}
G(z,w) = 2 \frac{|z-w|^2}{(1+z\zb)(1+w \wb)} \log   \frac{|z-w|^2}{(1+z\zb)(1+w \wb)}  .
\end{equation}
It will be convenient to define
\begin{equation}\label{sG}
s(z,w) = \gammaflatwt D_w^2 G(z,w) .
\end{equation}
Note that while $G(z,w)$ is real and symmetric in its two arguments, $s(z,w)$ is not. One can show that $s(z,w)$ obeys the following equations
\begin{eqnarray}
\label{eqs1} & D_{\zb}^2 s(z,w) = 2\pi \delta^{(2)}(z-w) , \\
\label{eqs2} & D_{\wb}^2 \left[ \gammaflatw s(z,w)\right)]  = 2\pi \gammaflatw \delta^{(2)}(z-w) .
\end{eqnarray}
Note also that the action of two covariant derivatives on a scalar function is given by
\begin{equation}
D_z^2=\pa_z^2 -\Gamma^z_{zz}\pa_z = \gammaflat \pa_z \gammaflatt\pa_z \equiv  \gammaflat \pa_z \pa^{\zb} .
\end{equation}

Using the decomposition $\ep^{\pm}_{\mu\nu}=\ep^{\pm}_{\mu}\ep^{\pm}_{\nu}$ we have
\begin{equation}
f\cdot \ep^{\pm} = \frac{(p\cdot \ep^{\pm})^2}{p\cdot q} - \frac{1}{\omega_q} c\cdot \ep^{\pm} ,
\end{equation}
where $c \cdot \ep^{\pm} \equiv c^{\mu\nu} \ep^{\pm}_{\mu\nu}$.
The first term in the expression above can be evaluated using the explicit form of the polarization vectors \eqref{polarization}
\begin{equation}
\begin{aligned}
\frac{(p\cdot \ep^+)^2}{p\cdot \hat{q}}  & = -s(z_0,w) , \\
\frac{(p\cdot \ep^-)^2}{p\cdot \hat{q}}  & = -\bar{s}(z_0,w) .
\end{aligned}
\end{equation}
We then have
\begin{equation}
\begin{aligned}
\Rfp &=
\frac{i}{2\pi}
\int d^2 w
\left[
(s(z_0,w)  + c \cdot \ep^+ )  C^-_{\wb\wb}(v_c,w,\wb)
+(\bar{s} (z_0,w)  + c \cdot \ep^- )   C^-_{ww}(v_c,w,\wb)
\right]
\\
&=
\frac{i}{2\pi}
\int d^2 w d^2 z
\big[
\delta^{(2)} (z-z_0 ) \big(  s(z,w)    C^-_{\wb\wb}(v_c,w,\wb)
+  \bar{s} (z,w)    C^-_{ww}(v_c,w,\wb) \big)
\\&
+
\delta^{(2)} (z-w ) \big( 
( c \cdot \ep^+)_z C^-_{\wb\wb}(v_c,w,\wb)
+ (c \cdot \ep^- )_z C^-_{ww}(v_c,w,\wb) \big)
\big] ,
\end{aligned}
\end{equation}
where, again, the subscript on $(c\cdot \ep^{\pm})_z$ is to indicate that now the expression in the parenthesis is evaluated at $(z,\zb)$, instead of $(w,\wb)$.
We now use equation \eqref{eqs1} to express the delta function in the second line of the equation above in terms of the function $s(z,w)$. After integration by parts we arrive at
\begin{equation}
\begin{aligned}
\Rfp
&=
\frac{i}{4\pi ^2}
\int d^2 w d^2 z \, 
\left(
s(z,w) \left[
2\pi \delta^{(2)}(z-z_0) +  \pa_{\zb} \pa^{z} (\gammaflat c\cdot \ep^+) 
\right]C^-_{\wb\wb}(v_c,w,\wb)
\right. \\& \left.
\qquad\qquad\qquad\quad \,\,
+\bar{s}(z,w) \left[
2\pi \delta^{(2)}(z-z_0) +  \pa_z \pa^{\zb} (\gammaflat c\cdot \ep^-) 
\right]C^-_{ww}(v_c,w,\wb)
\right) ,
\end{aligned}
\end{equation}
where we have omitted the subscript on $(\gammaflat c \cdot \ep^{\pm} )$ for brevity.

Finally, using eq.~\eqref{sG} to express the function $s(z,w)$ in terms of the Green's function and integrating by 
parts we arrive at 
\begin{equation}
\begin{aligned}
\Rfp 
&=
\frac{i}{ ( 2\pi)^2 }
\int d^2 w \gammaflatw d^2 z \gammaflat \, 
G(z,w) 
\big[
2\pi \gammaflatt \delta^{(2)}(z-z_0) D^2 \cdot C^-(v_c,w,\wb)
\\&
+ \pa^z\pa^z  (\gammaflat  c\cdot \ep^+) 
D^{\wb} D^{\wb}C^-_{\wb\wb}(v_c,w,\wb)
+
\pa^{\zb} \pa^{\zb} ( \gammaflat c\cdot \ep^-) 
D^w D^wC^-_{ww}(v_c,w,\wb)
\big]  ,
\end{aligned}
\end{equation}
where $D^2 \cdot C^- \equiv D^w D^w C^-_{ww}+D^{\wb}D^{\wb} C^-_{\wb\wb}$.

%If we further assume that $\int d^2 w \frac{C_{\wb\wb}}{(w-z)^2}=0$ {\color{red}(Here again we should discuss this assumption and whether it is valid when evaluated at $v\neq 0$)} we can arrive at
As in the gauge theory case, at this point we make the regularization choice $v_c = + \infty$ that isolates the contribution of zero momentum gravitons in the cloud. The boundary condition
\begin{equation}
\left[ D_{\zb}^2 C^-_{zz}-D_{z}^2 C^-_{\zb\zb}\right]_{\mI^-_{+}} =0
\end{equation}
then implies
\begin{equation}\label{gravCloud}
\begin{aligned}
\Rfp 
&=
\frac{i}{2\pi}
\int dv d^2 w \gammaflatw d^2 z \gammaflat \, 
G(z,w) 
\left[
T_{vv}(v,z,\zb)
-\bar{T}_{vv}(v,z,\zb)
\right]D^2 \cdot C^-(+ \infty,w,\wb) ,
\end{aligned}
\end{equation}
where
\begin{equation}
\begin{aligned}
T_{vv} (v,z,\zb) &= \delta(v-v_0) \gammaflatt \delta^{(2)} (z-z_0)  , \\
\bar{T}_{vv} (v,z,\zb) &=- \frac{1}{2\pi} \delta(v-v_0) \gammaflatt
\left[  \pa_{\zb}\pa^z  (\gammaflat  c\cdot \ep^+) 
+\pa_z \pa^{\zb} ( \gammaflat c\cdot \ep^-) \right] .
\end{aligned}
\end{equation}
$T_{vv} (v,z,\zb)$ is the classical stress energy tensor of the shockwave while 
$\bar{T}_{vv}(v,z,\zb)$ is interpreted as the supertranslation current, in accordance with ref.~\cite{Choi:2017bna}.

%%%%%%%%%%%%%%%%%%%%%%%%%%%%%%%%%%%%%%%%%%%%%%%%%%%%%%%%%%%%%%%%
%%%%%%%%%%%%%%%%%%%%%%%%%%%%%%%%%%%%%%%%%%%%%%%%%%%%%%%%%%%%%%%%
\section{Hawking Radiation of Dressed States}\label{BlackHole}
%%%%%%%%%%%%%%%%%%%%%%%%%%%%%%%%%%%%%%%%%%%%%%%%%%%%%%%%%%%%%%%%
%%%%%%%%%%%%%%%%%%%%%%%%%%%%%%%%%%%%%%%%%%%%%%%%%%%%%%%%%%%%%%%%

We now turn to study black hole horizons and the Hawking radiation. We start by analyzing the selection rule for S-matrix elements between Fock states, in the presence of a black hole horizon. Then we describe how the selection rule is modified once we consider dressed asymptotic states instead of Fock states. At this point we would like to emphasize that to linear order in the fluctuations around the background the asymptotic analysis of the previous sections applies to any null surface and in particular to a black hole horizon. The cloud operators, given by equations \eqref{photonCloud} and \eqref{gravCloud}, will take the same form, with the fields and charge densities evaluated at the horizon.

After discussing the selection rule we turn to a semi-classical analysis of the matter fields. We finally 
revisit Hawking's computation \cite{Hawking:1974sw,Hawking:1976ra} of particle creation near a black hole horizon, using the
formalism of  dressed asymptotic fields that we developed in the rest of the paper.  In this section we work in 
the limit $G\rightarrow 0$, $R_S=2GM=\mbox{constant}$, so that we consistently neglect back reaction of 
matter and Hawking's radiation on the background.

\subsection{Horizon selection rule}

Let us start by commenting on the selection rule for S-matrix elements in the presence of a black hole horizon. To be concrete, we consider a collapsing geometry generated by a null shockwave with initial energy $M$ sent at advanced time $v=0$.
The collapsing thin shell of matter will eventually form a Schwarzchild black hole of mass $M$. The resulting spacetime geometry is described by the Vaidya metric
\begin{equation}
ds^2 = - \left(   1- \frac{2M G \theta(v)}{r} \right) dv^2
+2 dv dr
+2r^2 \gamma_{z \zb} dz d \zb
\end{equation}
where $\theta$ denotes the step function
\begin{equation}
\theta (v) =
\begin{cases}
0 , \qquad v< 0  \\
1, \qquad v\geq 0 
\end{cases} .
\end{equation}
The horizon of the Vaidya metric is located at
\begin{equation}
r_H (v) =
\begin{cases}
0, \qquad & v<-4M \\
\frac{v}{2}+2M,  \qquad &-4M< v< 0  \\
2M, \qquad  & v\geq 0 
\end{cases}
\end{equation}
and is therefore stretched from $v= -4M$ to $v=\infty$.
\begin{figure}[tpb]
		\centering
\begin{tikzpicture}[scale=1]
\pgfmathsetmacro\myunit{4}
\pgfmathsetmacro\myunitShort{3.5}
\pgfmathsetmacro\myunitLong{4.5}
\pgfmathsetmacro\squareUnit{4.5}
\pgfmathsetmacro\squareUnitsmall{4}

\draw     (0,0)            node [left] {$i^0$} 
--++(45:\myunitShort)     coordinate (a)
node[pos=.5, above left]    {$\mI^+$} 
node [above]{$i^+$}
--++(-45:\myunitLong) coordinate (b)    
node[pos=.28, below,yshift=-2.5mm] {$\mH^+$} 
--++(45:\myunitLong)     coordinate (c)
--++(-45:\myunitShort)  coordinate (d) 
node[pos=.5, above, sloped] (u) {$u  $}     
node[pos=1, above, sloped] (dabove) {{\color{white}u}}
node[pos=0, above, sloped] (cabove) {{\color{white}u}}
--++(-135:2*\myunit)  coordinate (e)   
node[pos=.5, below, sloped] (v) {$v  $}     
node[pos=1, below, sloped] (eabove) {{\color{white}v}}
node[pos=0, below, sloped] (dabove2) {{\color{white}v}} 
node[pos=1, below]    {$i^-$}
node[pos=.5, below left,xshift=-58mm]    {$\mI^-$}
--cycle
;

\draw [decorate, decoration=zigzag] (a) -- node[above=6pt] {} (c)
;

\draw [dashed] (e) --++(45:\squareUnit)
--++(135:\squareUnit)
node[pos=.4, above, sloped] {\scriptsize $v=v_0$}    
--++(225:\squareUnit)
;

\draw [dashed] (e) --++(45:\squareUnitsmall)
--++(135:\squareUnitsmall)
node[pos=.45, below, sloped] {\scriptsize $v=0$}    
--++(225:\squareUnitsmall)
;

\draw [->] (dabove) -- (u)
[->] (u) -- (cabove)
 ;
 
 \draw [->] (eabove) -- (v)
 [->] (v) -- (dabove2)
 ;

\end{tikzpicture}
	\caption{Penrose diagram of the Vaidya black hole formed by a collapsing shell of a null shockwave at $v=0$. Another null shockwave is sent at a later advanced time $v=v_0$. Spacelike infinity is denoted by $i^0$ while $i^+/i^-$ is the 
	future/past timelike infinity.}
\label{fig:Vaidya}
\end{figure}

Incoming (outgoing) states are characterized by charges defined at past (future) null infinity
\begin{equation}
\begin{aligned}
Q^{\mI^-}_{\ep} &= \frac{1}{e^2}  \int _{\mI^- _+} d^2w \gammaflatw \ep F_{rv} ,\\
Q^{\mI^+}_{\ep} &= \frac{1}{e^2}  \int _{\mI^+ _-} d^2w \gammaflatw  \ep F_{ru} .
\end{aligned}
\end{equation}
Horizon states are similarly characterized by the charge
\begin{equation}
Q^{\mH}_{\ep} = \frac{1}{e^2}  \int _{\mH^+ _+} d^2w \gammaflatw \ep F_{rv} .
\end{equation}
The conservation law of these charges implies
\begin{equation}\label{conservationLaw}
\bra{\text{out}}
Q^+_{\ep} \mS
-
\mS Q^-_{\ep}
\ket{\text{in}}
=0 ,
\end{equation}
where $Q^+_{\ep}$ is the sum of the two charges
\begin{equation}
Q^+_{\ep} = Q^{\mI^+}_{\ep}  + Q^{\mH}_{\ep} .
\end{equation}

  We can rewrite the charges as integrals over the boundary of spacetime by extending
  the gauge parameter $\ep (z,\zb)$ along a null generator of the
  boundary~\cite{Hawking:2016msc}. A natural choice that makes the connection with
  soft quanta explicit is to keep the gauge parameter constant along a null
  generator of the (null) Cauchy surface.
  On $\mI^-$ this means that it does not depend on the
  advanced null coordinate $v$. In a similar way, the gauge parameter on $\mI^+$ is
  taken to be independent of the retarded null coordinate $u$.
  In the presence of a black hole horizon, however, the null
  Cauchy surface is not made of $\mI^+$ alone, but rather of its union with the
  horizon $\mI^+ \cup \mH^+$. This surface is the limit of a spacelike Cauchy
  surface which, by assumption, exhibits no singularity in the limit at $i^+$.
  A radial generator of the Cauchy surface then becomes a null generator of
  $\mI^+ \cup \mH^+$ which starts as the null generator $u$ of $\mI^+$, 
  ``turns at the corner'' at $i^+$, and continues as the null generator $v$ of
  $\mH^+$. The gauge parameter is then extended by taking it to be constant along this
  generator, so it takes the the same value at {\em equal} values of the
  angular coordinates
  on the horizon and at future null infinity $\mI^+$.

The soft parts of each of the charges are given by 
\begin{equation}
\begin{aligned}
N_z^- &=  \int_{-\infty}^{\infty} dv \, F_{vz}^{\mI^-}  ,\\
N_z^+ &=  \int_{-\infty}^{\infty} du \, F_{uz}^{\mI^+}  ,\\
N_z^{\mH^+}  &= \int_{-4M}^{\infty} dv \, ( F_{vz}^{\mH^{+}}  +\frac{1}{2} \theta(-v)F^{\mH^{+}} _{rz}  )  .
\end{aligned}
\end{equation}
If we choose $\ep(w,\wb) = \frac{1}{z-w}$, the conservation law can be written as
\begin{equation}\label{conservationLawHorizon}
\bra{\text{out}}
(N^+_{z}+ N_z^{\mH^+} ) \mS
-
\mS N^-_{z}
\ket{\text{in}}
=
\Omega^{\text{soft}}_z
\bra{\text{out}}
\mS
\ket{\text{in}} ,
\end{equation}
where
\begin{equation}
\Omega^{\text{soft}}=
\Omega^{\text{soft}-}_z 
-\Omega^{\text{soft}+}_z 
-\Omega^{\text{soft,}\mH^+}_z  .
\end{equation}
The soft factors are expressed in terms of the asymptotic particles coming from $\mI^-$ or going to $\mI^+$ and $\mH^+$
\begin{equation}\label{softFactors}
\begin{aligned}
\Omega^{\text{soft}-}_z &= \frac{e^2}{4\pi} \sum_{k \in \text{in}} \frac{Q_k}{z-z_k} , \\
\Omega^{\text{soft}+}_z &= \frac{e^2}{4\pi} \sum_{k \in \text{out}} \frac{Q_k}{z-z_k} , \\
\Omega^{\text{soft,}\mH^+}_z &= \frac{e^2}{4\pi} \sum_{k \in \text{horizon}} \frac{Q_k}{z-z_k} .
\end{aligned}
\end{equation}

Incoming, outgoing and horizon states are characterized by their eigenvalues under the corresponding soft charges 
\begin{equation}
\begin{aligned}
N_z ^- \ket{N_z^{\text{in}}} &= N_z^{\text{in}}(z,\zb)\ket{N_z^{\text{in}}} , \\
N_z ^+ \ket{N_z^{\text{out}}} &= N_z^{\text{out}}(z,\zb)\ket{N_z^{\text{out}}} , \\
N_z ^{\mH^+} \ket{N_z^{\text{H}}} &= N_z^{\text{H}}(z,\zb)\ket{N_z^{\text{H}}}  .
\end{aligned}
\end{equation}
The conservation law \eqref{conservationLawHorizon} then reduces to
\begin{equation}\label{conred}
\left(
N^{\text{out}}_{z}+ N_z^{\text{H}} -N^{\text{in}}_{z}
\right)
\bra{\text{out}}
\mS
\ket{\text{in}}
=
\Omega^{\text{soft}}_z
\bra{\text{out}}
\mS
\ket{\text{in}}
\end{equation}
and it implies that either
\begin{equation}\label{cond1}
N^{\text{out}}_{z}+ N_z^{\text{H}} -N^{\text{in}}_{z}= \Omega^{\text{soft}}_z ,
\end{equation}
or the S-matrix element vanishes
\begin{equation}
\bra{\text{out}}
\mS
\ket{\text{in}}=0  .
\end{equation}
Equation \eqref{cond1} cannot be satisfied in general and therefore we conclude that S-matrix elements between Fock states vanish.

We now turn to consider dressed asymptotic states instead of Fock states.
Using the canonical commutation relation
\begin{equation}\label{commrel}
\left[
A_{w} (v,w,\wb)
,
F_{v'\zb}(v',z,\zb)
\right]
=
\frac{ie^2}{2} \delta(v-v') \delta^{(2)}(w-z) 
\end{equation}
we see that the dressing shifts the action of $F_{vz}$ on states by
\begin{equation}
\begin{aligned}
\left[ F_{vz} (v,z,\zb)  , \Rfp \right] &=-\frac{e^2 \delta(v-v_c)}{4\pi} \int dv' d^2 z' \gamma_{z'\zb '}\frac{j_{v'}(v',z',\zb ') - \jb_{v'}  (v',z',\zb ')}{z'-z}
\\
&= -\frac{Q_0 e^2 \delta(v-v_c)}{4\pi (z-z_0)} +\frac{e^2 \delta(v-v_c)}{4\pi} \int d^2z' \frac{\jb_{v'}(z',\zb')}{z'-z} ,
\end{aligned} 
\end{equation}
where we used the expression \eqref{photonCloud} for the photons' cloud operator.
The delta function in the above expression is centered around $v=v_c$ because we evaluate the cloud operator \eqref{photonCloud} at $v_c$. The choice $v_c=+\infty$ isolates the contributions of the soft modes but 
 the action of the soft charge on incoming multi-particle state given by the $v_c$-independent formula
\begin{equation}
N_z ^- \ketd[N_z^{\text{in}}] =
\left[
N_z ^{\text{in}}    
-
\sum_{j \in \text{in}}\left(
\frac{Q_j e^2}{4\pi(z-z_j)}
-\frac{e^2}{4\pi} \int d^2z' \frac{\jb_{v'}^{(j)}(z',\zb')}{z'-z_j}
\right)
\right]
\ketd[N_z^{\text{in}}] .
\end{equation}
Similarly, for outgoing and horizon multi-particle states we have, respectively
\begin{equation}
\begin{aligned}
\brad[N_z^{\text{out}}] N_z ^+ &= 
\brad[N_z^{\text{out}}]
\left[
N_z ^{\text{out}}    
-
\sum_{i \in \text{out}}\left(
\frac{Q_i e^2}{4\pi(z-z_i)}
-\frac{e^2}{4\pi} \int d^2z' \frac{\jb_{v'}^{(i)}(z',\zb')}{z'-z_i}
\right)
\right] ,
\\
\brad[N_z^{\text{out}}] N_z ^{\mH^+} &= 
\brad[N_z^{\text{out}}]
\left[
N_z ^{\text{H}}    
-
\sum_{k \in \text{Horizon}}\left(
\frac{Q_k e^2}{4\pi(z-z_k)}
-\frac{e^2}{4\pi} \int d^2z' \frac{\jb_{v'}^{(k)}(z',\zb')}{z'-z_k}
\right)
\right] .
\end{aligned}
\end{equation}
Thanks to \eqref{conred} conservation law \eqref{conservationLaw} then becomes 
\bea
&&\left[
\left( N_z ^{\text{out}} +N_z ^{\text{H}}     -N_z ^{\text{in}}  \right)
\right. \\ && \left.
\nonumber
-\frac{ e^2}{4\pi} \int d^2z' 
\left(
\sum_{i \in \text{out}} \frac{\jb_{v'}^{(i)}(z',\zb')}{z'-z_i} 
+\sum_{k \in \text{Horizon}} \frac{ \jb_{v'}^{(k)}(z',\zb')}{z'-z_k} 
- \sum_{j \in \text{in}} \frac{\jb_{v'}^{(j)}(z',\zb')}{z'-z_j}
\right)
\right]
\bra{\text{out}}_{\text{dressed}}
\mS
\ket{\text{in}}_{\text{dressed}}
=0 .
\eea
a
As was explained in \cite{Gabai:2016kuf}, the integrand inside the $z'$-integral above should vanish\footnote{The authors of \cite{Gabai:2016kuf} made a different choice of the transformation parameter $\ep (w,\wb)$.}, and therefore the conservation law reduces to
\begin{equation}
\left(
N_z ^{\text{out}}  +N_z ^{\text{H}}    -N_z ^{\text{in}}    
\right)
\bra{\text{out}}_{\text{dressed}}
\mS
\ket{\text{in}}_{\text{dressed}}
=0 .
\end{equation}
It implies that the S-matrix element between dressed asymptotic states does not vanish when the soft charge is conserved
\begin{equation}\label{selectionLGT}
\bra{\text{out}}_{\text{dressed}}
\mS
\ket{\text{in}} _{\text{dressed}} \neq 0 
\qquad
\text{for}
\qquad
N_z ^{\text{out}}  +N_z ^{\text{H}}    =N_z ^{\text{in}}     .
\end{equation}
Notice that the LGT current $\jb_v$ do not affect the selection rule \eqref{selectionLGT}.

\subsection{The action of the cloud operator}

We now turn to a semi-classical analysis of matter fields.
We start by looking at the action of the cloud operator on an incoming massless scalar.
The effect of the photon cloud operator on matter fields is rather trivial, since it commutes with all of them up to a phase
proportional to the charge --see eq.~\eqref{m6}. However, since the graviton cloud operator contains the stress-energy tensor it acts in a non-trivial way on any field.
The dressed scalar field is given by
\begin{equation}
\phiinhat \equiv e^{-\Rfp} \phiin e^{\Rfp} ,
\end{equation}
where the graviton dressing operator $\Rfp$ is given by \eqref{gravCloud}.

Let us summarize our notations for the scalar field. First of all, it can be expanded around null infinity as follows
\begin{equation}
\Phi(r,v,z,\zb) = \frac{1}{r} \phi (v,z,\zb) + \mO(r^{-2}) .
\end{equation}
Its Fourier decomposition is given by
\begin{equation}
\Phi (\vec{x}) = \int \frac{d^3p}{(2\pi)^3 2\omega_p} \left[
d(p) e^{ip\cdot x} + d^{\dagger}(p) e^{-ip\cdot x}
\right] ,
\end{equation}
with the Fourier components obeying the canonical commutation relations
\begin{equation}
\left[d(p), d^{\dagger}(p') \right] = (2\pi)^3 2 \omega_p \delta^{(3)}(\vec{p} -\vec{p'}) .
\end{equation}
Using the asymptotic expansion for $\exp(\pm i p \cdot x)$
 (see for example appendix A of \cite{Gabai:2016kuf}) we then have
\begin{equation}
\phiin(v,z,\zb) =-i
\int \frac{d\omegap}{8\pi ^2}\left[
d(p) e^{-i \omegap v}
-d^{\dagger}(p)  e^{i \omegap v}
\right] .
\end{equation}
The field therefore obeys the following commutation relation
\begin{equation}
\left[\phiin(v,z,\zb) , \pa_v \phiin(v',z',\zb ')\right] =\frac{i}{4} \gammaflatt \delta(v-v') \delta^{(2)} (z-z') .
\end{equation}

The stress tensor is given by
\begin{equation}
T_{vv} = - \frac{1}{2} (\pa_v \phiin)^2 ,
\end{equation}
thus we have
\begin{equation}
\left[\phiin(v,z,\zb) , T_{vv} (v',z',\zb ')\right] =- \frac{i}{4} \gammaflatt \pa_v \phiin(v,z,\zb) \delta(v-v') \delta^{(2)} (z-z') .
\end{equation}
One can then compute
\begin{equation}
\begin{aligned}
\left[\phiin(v,z,\zb) ,  \Rfp \right] &=
-\gammaflat \pa_v \phiin(v,z,\zb)
\int d^2 w \gammaflatw
G(z,w)
D^2 \cdot C^-(\pm \infty,z,\zb)
\\
&=
-\gammaflat \pa_v \phiin(v,z,\zb)
\int d^2 w \left[
s(z,w) C^-_{\wb\wb} (\pm \infty,z,\zb)+\bar{s}(z,w) C^-_{ww}(\pm \infty,z,\zb)
\right] ,
\end{aligned}
\end{equation}
where in the second line we used integration by parts. The boundary conditions further implies that
\begin{equation}
C^-_{zz} (\pm\infty , z,\zb)= -2 D^2 _z C^-(z,\zb) .
\end{equation}
Integrating by parts and using \eqref{eqs2} we arrive at
\begin{equation}
\left[\phiin(v,z,\zb) , \Rfp \right] =
-C^-(z,\zb) \pa_v \phiin(v,z,\zb) .
\end{equation}

Now we can expand the dressed asymptotic state  using the Baker-Campbell-Hausdorff formula as follows
\begin{equation}
\begin{aligned}
\phiinhat(v,z,\zb) &= e^{-\Rfp} \phiin(v,z,\zb) e^{\Rfp} \\
&  = \phiin(v,z,\zb) + [\phiin(v,z,\zb),\Rfp] +\frac{1}{2!}[[\phiin(v,z,\zb),\Rfp],\Rfp]+\dots
\end{aligned}
\end{equation}
The $n$-th commutator of $\phiin$ with $\Rfp$ is given by
\begin{equation}
\left[ \left[ \left[\phiin(v,z,\zb) ,  \Rfp \right] ,\Rfp \right],\dots \right]_n=
-C^-(z,\zb) \pa_v^n \phiin(v,z,\zb) .
\end{equation}

We have therefore showed that 
\begin{equation}\label{asyState}
\begin{aligned}
\phiinhat(v,z,\zb) & \equiv e^{-\Rfp} \phiin(v,z,\zb) e^{\Rfp} \\
&=  \sum_{n=0}^{\infty} (-C^-(z,\zb))^n \pa_v ^n\phiin(v,z,\zb) =  \phiin(v-C^-(z,\zb),z,\zb) .
\end{aligned}
\end{equation}
We see that the dressing of the incoming scalar state is equivalent to a shift in the null direction.
This result was derived in \cite{Bousso:2017dny} using different methods.
Note that, in particular, the supertranslation current $\bar{T}_{vv}$ does not affect the dressed scalar state.

\subsection{Particle creation in a black hole background}

To adjust for the normalization most commonly used in the literature on Hawking radiation, we rescale
\begin{equation}
d(\omega) = i \frac{8\pi^2 }{\sqrt{2\pi \omega}} a(\omega) .
\end{equation}
Then we have the following expansion of incoming and outgoing modes
\begin{equation}
\begin{aligned}
\phiin(v,z,\zb) &= \int _{0}^{\infty} \frac{d\omega}{\sqrt{2\pi \omega}}\left(
a_{\omega} e^{-i \omega v}
+a^{\dagger}_{\omega} e^{i \omega v}
\right) , \\
\phiout(u,z,\zb) &= \int _{0}^{\infty} \frac{d\omega}{\sqrt{2\pi \omega}}\left(
b_{\omega} e^{i \omega u}
+b^{\dagger}_{\omega} e^{-i \omega u}
\right) .
\end{aligned}
\end{equation}
Hawking \cite{Hawking:1974sw,Hawking:1976ra} found the relation between the incoming and outgoing modes
\begin{equation}\label{relations}
\begin{aligned}
b_{\omega} &= \int_0 ^{\infty} d\omega ' \left( \alpha_{\omega \omega '} a_{\omega '} + \beta_{\omega \omega '} a_{\omega '}^{\dagger}  \right) , \\
b_{\omega}^{\dagger} &= \int_0 ^{\infty} d\omega ' \left( \alpha_{\omega \omega '}^* a_{\omega '}^{\dagger} + \beta_{\omega \omega '}^*  a_{\omega '}  \right)  ,
\end{aligned}
\end{equation}
where the explicit form of the universal, late-time Bogolyubov coefficients is
\begin{equation}\label{bogoCo}
\begin{aligned}
\alpha_{\omega \omega '} &=  \frac{t_{\omega} }{2\pi} e^{i (\omega -\omega ' )v_0} \sqrt{\frac{\omega ' }{\omega}} \Gamma(1-i \kappa ^{-1}\omega) (-i \omega ') ^{-1+i \kappa^{-1} \omega}  , \\
\beta_{\omega \omega '} &= - i \alpha_{\omega (-\omega ')}  .
\end{aligned}
\end{equation}

We would like to consider now asymptotic states dressed with a graviton cloud \eqref{asyState} 
\begin{equation}
\begin{aligned}
\phiinhat(v,z,\zb) &=\phiin(v-C^-(z,\zb),z,\zb) = \int _{-\infty}^{\infty} \frac{d\omega}{\sqrt{2\pi \omega}}\left(
\hat{a}_{\omega} e^{-i \omega v}
+\hat{a}^{\dagger}_{\omega} e^{i \omega v}
\right)  , \\
\phiouthat(u,z,\zb) &=\phiout(u-C(z,\zb),z,\zb) = \int _{-\infty}^{\infty} \frac{d\omega}{\sqrt{2\pi \omega}}\left(
\hat{b}_{\omega} e^{i \omega u}
+\hat{b}^{\dagger}_{\omega} e^{-i \omega u}
\right) ,
\end{aligned}
\end{equation}
where
\begin{equation}
\begin{aligned}
\hat{a}_{\omega} &= a_{\omega} e^{+i \omega C^-(z,\zb)} , \\
\hat{b}_{\omega} &= b_{\omega} e^{-i \omega C(z,\zb)} . 
\end{aligned}
\end{equation}
Note that the boundary conditions on the boundary graviton at $i^0$ impose $C(z,\zb)=-C^-(z,\zb)$, but we will keep them 
independent for now.

To compute the effect of soft hair on the process it suffices to recall that eq.~\eqref{m9} in the Introduction implies that 
dressed variables evolve independently of soft variables. It is straightforward to check that eq.~\eqref{m9} applies also 
to the Vaydia metric. In fact the fields defined at $\mathcal{I}^\pm$ are independent of {\em all} variables defined at the
horizon by locality, so the derivation given in the Introduction can be carried out word by word also in the present case. 
Since dressed variables do not interact with soft variables, while their interactions with radiative variables are proportional to $G$, they do interact only with the background in the limit $G\rightarrow 0$,  $R_S=2GM=\mbox{constant}$. 
This means that the standard derivation of Hawking follows through unmodified for the dressed variables; namely:
\begin{equation}\label{dressedRelations}
\begin{aligned}
\hat{b}_{\omega} &= \int_0 ^{\infty} d\omega ' \left( \alpha_{\omega \omega '} \hat{a}_{\omega '} + \beta_{\omega \omega '} \hat{a}_{\omega '}^{\dagger}  \right) , \\
\hat{b}_{\omega}^{\dagger} &= \int_0 ^{\infty} d\omega ' \left( \alpha_{\omega \omega '}^* \hat{a}_{\omega '}^{\dagger} + \beta_{\omega \omega '}^*  \hat{a}_{\omega '}  \right) .
\end{aligned}
\end{equation}
The coefficients $\alpha_{\omega \omega '}$ and $\beta_{\omega \omega '}$ are exactly those given in~\eqref{bogoCo}. 
This implies the following relation for the \emph{undressed} operators
\begin{equation}
\begin{aligned}
b_{\omega} &= \int_0 ^{\infty} d\omega ' \left( \hat{\alpha}_{\omega \omega '} a_{\omega '} + \hat{\beta}_{\omega \omega '} a_{\omega '}^{\dagger}  \right) , \\
b_{\omega}^{\dagger} &= \int_0 ^{\infty} d\omega ' \left(  \hat{\alpha}_{\omega \omega '}^* a_{\omega '}^{\dagger} + \hat{\beta}_{\omega \omega '}^*  a_{\omega '}  \right)  ,
\end{aligned}
\end{equation}
where
\begin{equation}\label{dressedCo}
\begin{aligned}
\hat{\alpha}_{\omega \omega '} &=  \alpha_{\omega \omega '}  e^{i \left[  \omega C(z,\zb) + \omega ' C^-(z,\zb) \right]  } , \\
\hat{\beta}_{\omega \omega '} &= \beta_ {\omega \omega '}  e^{i \left[  \omega C(z,\zb)  - \omega ' C^-(z,\zb) \right] } .
\end{aligned}
\end{equation}

The flux of outgoing particles is given by
\begin{equation}
\left\langle n  \right\rangle =\langle b^{\dagger}_{\omega}b_{\omega}  \rangle .
\end{equation}
More generally, one could calculate the correlation function
\begin{equation}
\begin{aligned}
\langle b^{\dagger}_{\omega_1}b_{\omega_2}    \rangle 
&=
\int d \omega '  \hat{\beta}_{\omega_1 \omega '} ^* \hat{\beta}_{\omega_2 \omega '} \\
&=
e^{ -i(\omega_1 - \omega _2 )C(z,\zb)}  \int d \omega '  \beta_{\omega_1 \omega '} ^* \beta_{\omega_2 \omega '} .
\end{aligned}
\end{equation}
Following Hawking's derivation \cite{Hawking:1974sw,Hawking:1976ra} the last integral can be evaluated to give
\begin{equation}
\begin{aligned}
\langle b^{\dagger}_{\omega_1}b_{\omega_2}    \rangle 
&=
e^{-i(\omega_1 - \omega _2 )C(z,\zb)}  \langle b^{\dagger}_{\omega_1}b_{\omega_2}    \rangle  _{\text{Hawking}}\\
&=
e^{-i(\omega_1 - \omega _2 )C(z,\zb)}  |t_{\omega}|^2 (e^{2\pi \omega \kappa^{-1}}-1)^{-1} \delta(\omega_1-\omega_2) .
\end{aligned}
\end{equation}
We see that the dressing operators introduce a phase factor $e^{-i(\omega_1 - \omega _2 )C(z,\zb)}$ in the result for the flux of outgoing particles, compared to the original derivation of Hawking. However, this phase disaapears as a result of the delta function $\delta (\omega_1 - \omega _2)$.
Higher point functions will also contain a phase factors 
\begin{equation}
\left\langle n ^m  \right\rangle =
e^{-i m (\omega_1 - \omega _2 )C(z,\zb)} \left\langle n ^m  \right\rangle_{\text{Hawking}} ,
\end{equation}
that will vanish in a similar manner due to the delta function.

Let us have another look at the effect of the phase factor. Using the boundary condition $C(z,\zb)=-C^-(z,\zb)$ we see that the dressed Bogolyubov coefficients are related to the undressed ones by
\begin{equation}
\begin{aligned}
\hat{\alpha}_{\omega \omega '} &= \left[ \alpha_{\omega \omega '} \right] _{v_0 \rightarrow v_0 + C(z,\zb)} ,\\
\hat{\beta}_{\omega \omega '} &= \left[ \beta_{\omega \omega '} \right] _{v_0 \rightarrow v_0 + C(z,\zb)} .
\end{aligned}
\end{equation}
Namely, the effect of the dressing is to shift $v_0$ by an amount $C(z,\zb)$.

We end this section with a few remarks on the result \eqref{dressedCo}. First, we would like to emphasize that the effect of 
the dressing on the Bogolyubov coefficients \eqref{dressedCo} is independent of the details of the gravitational collapse. 
Namely, corrections due the the details of collapse will only affect the undressed Bogolyubov coefficients 
$\alpha_{\omega \omega '}, \beta_{\omega \omega '}$, while the dependence on the dressing in \eqref{dressedCo} will 
remain unchanged. We also note that the phase factor does not appear in any observable. As we have seen, it vanishes in 
the correlator $\langle b^{\dagger}_{\omega_1}b_{\omega_2}    \rangle$ due to the delta function 
$\delta(\omega_1-\omega_2)$. Other correlation functions, like $\langle b_{\omega_1} b_{\omega_2} \rangle \sim \delta(\omega_1 + \omega _2)$, which could potentially depend on the dressing's phase factor, vanish identically due the 
delta function since both $\omega_1,\omega _2$ are 
positive. These results imply in particular that correlation functions of the form 
$\langle \phiout(u_1, z_1 ,\zb _1) \phiout(u_2, z_2 ,\zb _2) \rangle$ will not depend on the dressing's phase factor. 
We therefore conclude that the dressing is not observable with local operators defined on $\mathcal{I}^+$,
for a gravitational collapse and particle creation near a black hole horizon in the limit 
$G\rightarrow 0$, $R_S=\text{constant}$.

%%%%%%%%%%%%%%%%%%%%%%%%%%%%%%%%%%%%%%%%%%%%%%%%%%%%%%%%%%%%%%%%
%%%%%%%%%%%%%%%%%%%%%%%%%%%%%%%%%%%%%%%%%%%%%%%%%%%%%%%%%%%%%%%%
%\section{Conclusions}
%%%%%%%%%%%%%%%%%%%%%%%%%%%%%%%%%%%%%%%%%%%%%%%%%%%%%%%%%%%%%%%%
%%%%%%%%%%%%%%%%%%%%%%%%%%%%%%%%%%%%%%%%%%%%%%%%%%%%%%%%%%%%%%%%

%%%%%%%%%%%%%%%%%%%%%%%%%%%%%%%%%%%%%%%%%%%%%%%%%%%%%%%%%%%%%%%%
%%%%%%%%%%%%%%%%%%%%%%%%%%%%%%%%%%%%%%%%%%%%%%%%%%%%%%%%%%%%%%%%
\acknowledgments
The work of UK and MP is supported in part by NSF through grant PHY-1620039.

%%%%%%%%%%%%%%%%%%%%%%%%%%%%%%%%%%%%%%%%%%%%%%%%%%%%%%%%%%%%%%%%
%%%%%%%%%%%%%%%%%%%%%%%%%%%%%%%%%%%%%%%%%%%%%%%%%%%%%%%%%%%%%%%%

%%%%%%%%%%%%%%%%%%%%%%%%%%%%%%%%%%%%%%%%%%%%%%%%%%%%%%%%%%%%%%%%
%%%%%%%%%%%%%%%%%%%%%%%%%%%%%%%%%%%%%%%%%%%%%%%%%%%%%%%%%%%%%%%%
\appendix
%%%%%%%%%%%%%%%%%%%%%%%%%%%%%%%%%%%%%%%%%%%%%%%%%%%%%%%%%%%%%%%%
%%%%%%%%%%%%%%%%%%%%%%%%%%%%%%%%%%%%%%%%%%%%%%%%%%%%%%%%%%%%%%%%

%%%%%%%%%%%%%%%%%%%%%%%%%%%%%%%%%%%%%%%%%%%%%%%%%%%%%%%%%%%%%%%%
%%%%%%%%%%%%%%%%%%%%%%%%%%%%%%%%%%%%%%%%%%%%%%%%%%%%%%%%%%%%%%%%
%%%%%%BIBLIOGRAPHY 
%%%%%%%%%%%%%%%%%%%%%%%%%%%%%%%%%%%%%%%%%%%%%%%%%%%%%%%%%%%%%%%%
%%%%%%%%%%%%%%%%%%%%%%%%%%%%%%%%%%%%%%%%%%%%%%%%%%%%%%%%%%%%%%%%

\bibliographystyle{ssg}

%\nocite{*}

\bibliography{AsymptoticBH.bib}

\end{document}